\documentclass[preprint,12pt,authoryear]{elsarticle}

\usepackage{amssymb}
\usepackage{lscape}
\usepackage{rotating}
\usepackage{amsmath}
\usepackage[utf8]{inputenc}
\newcommand\shorttitle{IC\,5146 complex and surroundings}

\journal{ASSc}

\begin{document}

\begin{frontmatter}



\title{The IC\,5146 star forming complex and its surroundings with 2MASS, WISE and Spitzer}


%\shorttitle{IC\,5146 complex and surroundings\\} 
%\authors{Nunes, Bonatto and Bica}

\author{N.A.Nunes $^1$, C.Bonatto,$^1$. and E.Bica,$^1$}

\address{ $^1$Departamento de Astronomia, Universidade Federal do Rio Grande do Sul, Av. Bento
Gon\c{c}alves, 9500\\
Porto Alegre 91501-970, RS, Brazil\\}

\begin{abstract}
Throughout the last decade sensitive infrared observations obtained by the Spitzer Space Telescope significantly increased the known population of YSOs associated with nearby molecular clouds. With such a census recent studies have characterized pre-main sequence stars (PMS) and determined parameters from different wavelengths. Given the restricted Spitzer coverage of some of these clouds, relative to their extended regions, these YSO populations may represent a limited view of star formation in these regions. We are taking advantage of mid-infrared observations from the NASA Wide Field Infrared Survey Explorer (WISE), which provides an all sky view and therefore full coverage of the nearby clouds, to assess the degree to which their currently known YSO population may be representative of a  more complete population. We extend the well established classification method of the Spitzer Legacy teams to archived WISE observations. We have adopted 2MASS photometry as a “standard catalogue” for comparisons.  Besides the massive embedded cluster IC\,5146 we provide a multiband view of five new embedded clusters in its surroundings that we discovered with WISE. In short, the analysis involves the following for the presently studied cluster sample: {\em (i)} extraction of 2MASS/WISE/Spitzer photometry in a wide circular region; {\em (ii)} field-star decontamination to enhance the intrinsic Colour-magnitude diagram (CMD) morphology (essential for a proper derivation of reddening, age, and distance from the Sun); and (iii) construction of Colour-magnitude filters, for more contrasted stellar radial density profiles (RDPs).

\end{abstract}

\end{frontmatter}


%\section{}
%\label{}

\section{Introduction}
\label{intro}

Star formation occurs inside Giant Molecular Clouds (GMCs) and surveys of molecular gas in galaxies show that it is typically concentrated in large complexes or spiral arm segments having sizes up to a kiloparsec and masses up to $10^{7}$ $M_\odot$ (Solomon \& Sanders,1985 and Elmegreen, 1993). These complexes may contain several giant molecular clouds (GMCs) with sizes up to $100$ parsecs (pc) and masses up to $10^{6} M_\odot$. The GMCs in turn, contain much smaller scale structures that may be filamentary or clumpy on a wide range of scales (Blitz, 1993; Blitz \& Williams, 1999; Williams, Blitz \& McKee, 2000). The substructures found in GMCs range from massive clumps with sizes of several parsecs and thousands of solar masses, which may form entire clusters of stars, to small dense cloud cores with sizes of the order of $0.1$ pc and masses of the order of $1 M_\odot$ (Mendoza, 1985; Cernicharo,1991; Larson, 1994; Williams, Blitz \& McKee, 2000; Andre, Ward-Thompson \& Barsony, 2000; Visser, Richer \& Chandler, 2002). The internal structure of molecular clouds is partly hierarchical, consisting of smaller subunits within larger ones (Scalo,1990; Larson,1995; Simon, 1997; Stutzki et al., 1998; Elmegreen, 2000). In particular, the irregular boundaries of molecular clouds have fractal-like shapes resembling those of surfaces in turbulent flows, and this might mean that the shapes of molecular clouds are created by turbulence (Falgarone, Phillips \& Walker, 1991; Falgarone, Puget \& Perault, 1992).\\
Most molecular clouds form stars, but very inefficiently, typically turning only a few percent of their mass into stars before being dispersed. Despite the strong dominance of gravity over thermal pressure, this low efficiency has long been considered problematic, and implying that additional effects such as magnetic fields, angular momentum conservation or turbulence support these clouds in near-equilibrium against gravity and prevent a rapid collapse (Heiles et al., 1993; McKee et al., 1993).\\
\qquad The lifetimes and evolution of molecular clouds is provided by the ages of the associated newly formed stars and star clusters (Blaauw, 1991; Larson,1994; Elmegreen, 2000; Andre, Ward-Thompson \& Barsony, 2000; Hartmann, Ballesteros-Paredes \& Bergin, 2001). Very few GMCs are known that are not forming stars, and the most massive and dense ones as a rule contain newly formed stars. They also cannot survive for long after beginning to make stars, since the age span of the associated young stars and clusters is never more than $\sim 10$ Myr, about the dynamical or crossing time of a large GMC. Stars and clusters older than $10$ Myr do not appear to be associated with molecular gas (Leisawitz, Bash \& Thaddeus, 1989).\\
The youngest stars are associated with the denser parts of molecular clouds, and especially with the densest cloud cores that appear to be the direct progenitors of stars and stellar groupings (Mendoza,1985; Lada, Strom \& Myers, 1993; Williams, Blitz \& McKee, 2000; Andre,Ward-Thompson \& Barsony, 2000).\\
\qquad After several decades of study, the physical conditions in nearby star-forming molecular clouds are now fairly well understood, at least for the smaller nearby dark clouds that form mostly low-mass stars. However in the past $40$ years, infrared observations have revolutionized our understanding of star formation (Gutermuth et al., 2009). The emission excess  of young stars is well above that expected from reddened stellar photospheres and originates from the dusty circumstellar disks and envelopes surrounding young stars. For these reasons, colour-colour diagrams and CMDs in the infrared (IR) have proven to be excellent tools for identifying and classifying young stellar objects (Megeath et al. 2004; Harvey et al. 2008; Gutermuth et al. 2009).\\
\qquad We intend to address the physical mechanisms responsible for the formation of PMS stars in embedded star clusters  which is crucial for understanding their properties.\\
\qquad In Sect.~\ref{history} we discuss the IC\,5146 nebular complex to which a Streamer appears to be related, and contains a sample of 5 newly found embedded clusters (ECs). In Sect.~\ref{sample} we gather the data. In Sect.~\ref{CMDs} we discuss the general properties of the objects. In Sect. ~\ref{disc} we discuss the results. Finally, in Sect.~\ref{conclu} we provide the conclusion of this study.

\section{Overview of the IC 5146 nebular complex}
\label{history}

\begin{figure}
\resizebox{\hsize}{!}{\includegraphics{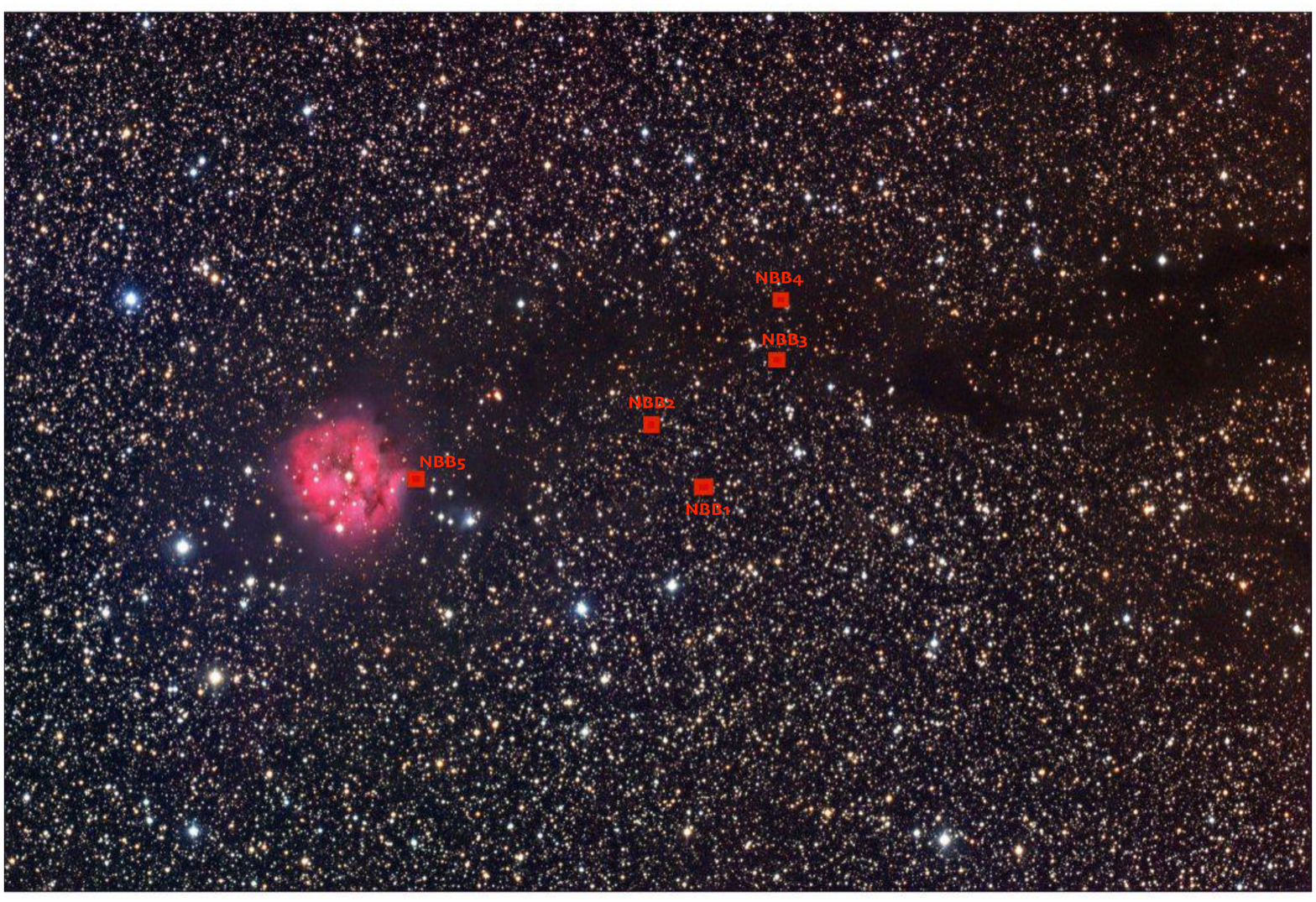}}
\caption{NOAO/AURA/NSF optical image from  Kitt Peak telescope (Field is $\sim$ $2\,^{\circ} \times 1.5\,^{\circ}$) of IC\,5146 (left) and the positions of the five new ECs (squares) along the Streamer. North to the top and east to the right. Image credit: Adam Block.}
\label{IC5146}
\end{figure}

\begin{figure*}
\begin{minipage}[b]{0.33\linewidth}
\includegraphics[width=\textwidth]{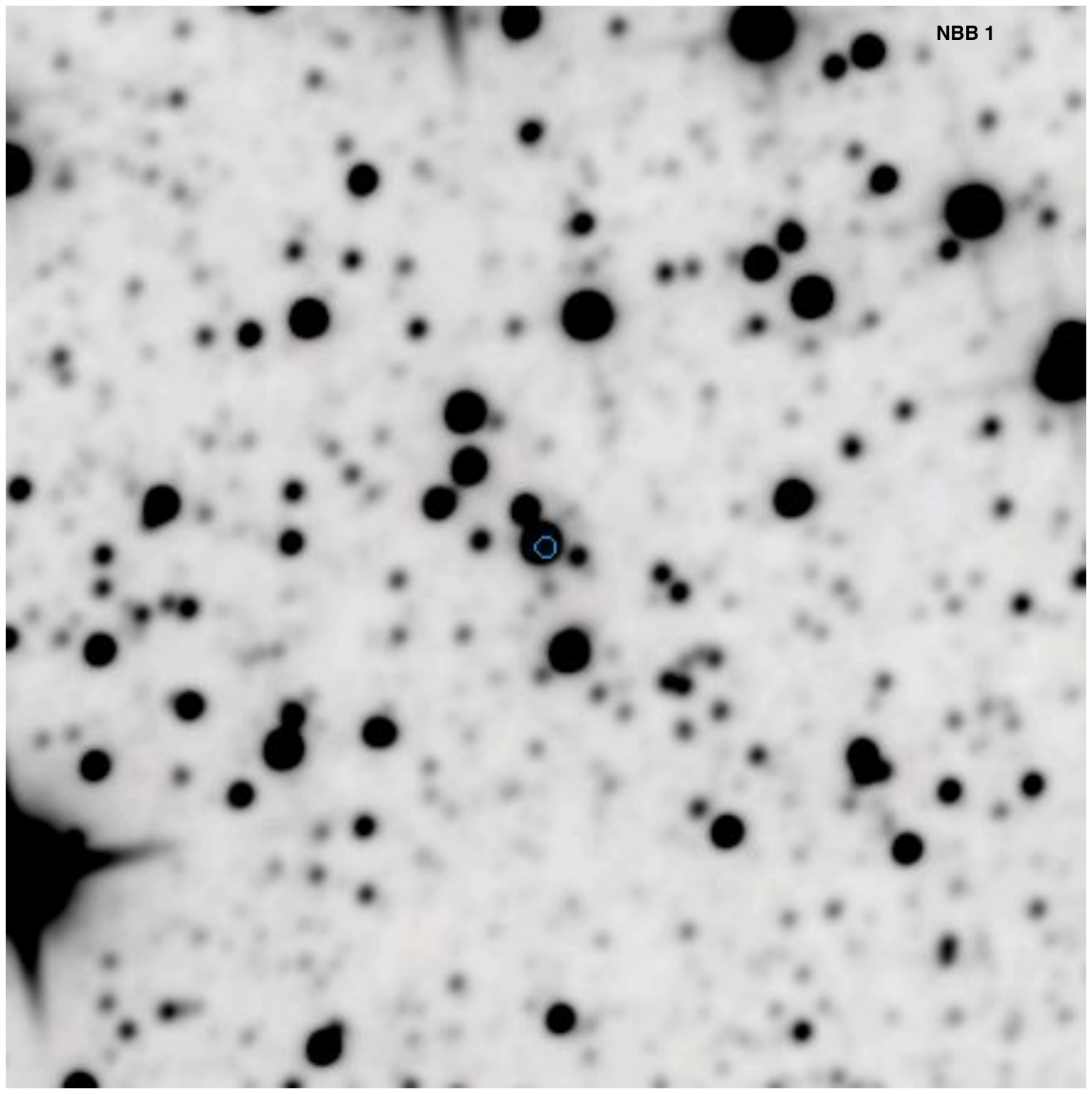}
\end{minipage}\hfill
\begin{minipage}[b]{0.33\linewidth}
\includegraphics[width=\textwidth]{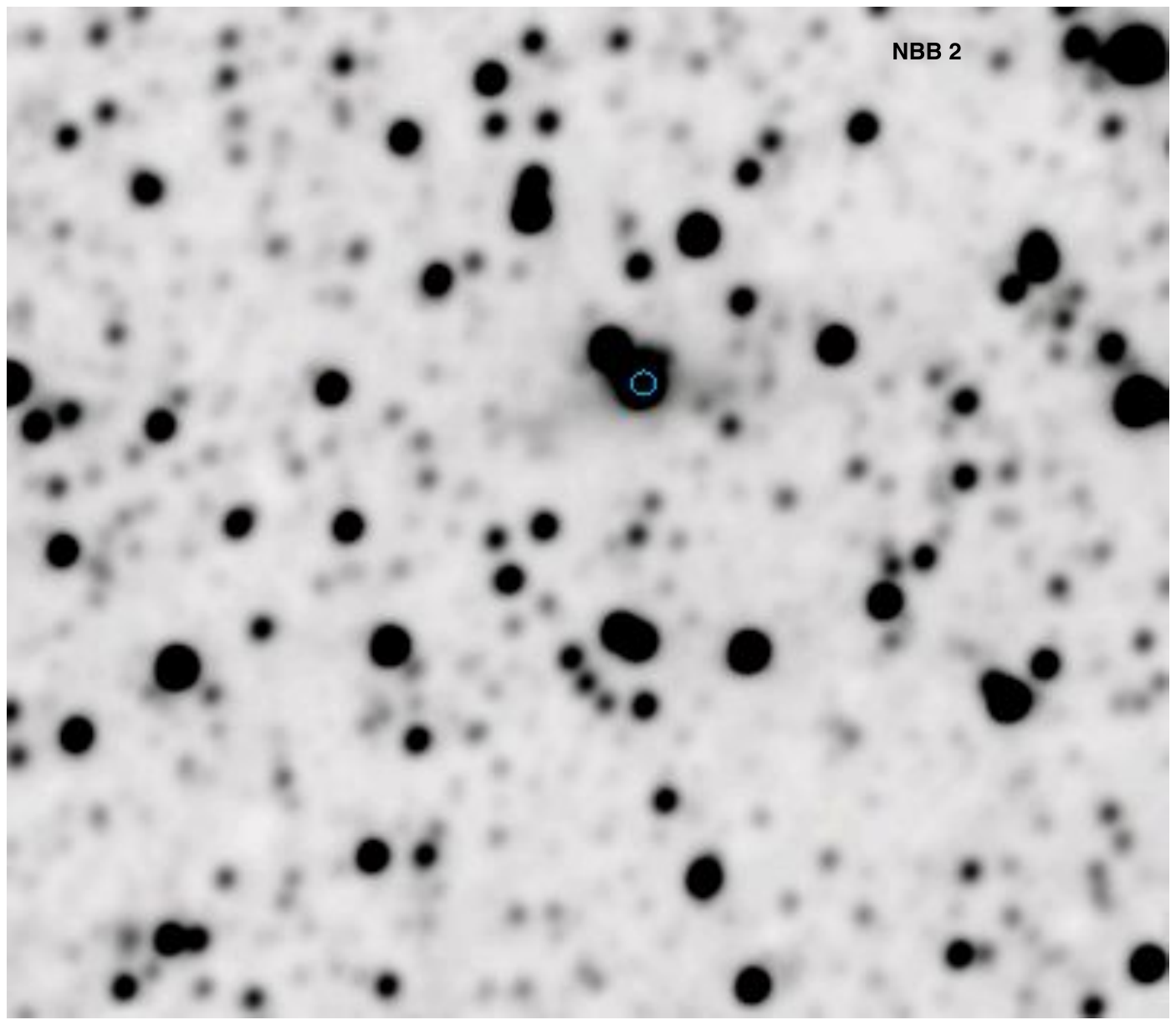}
\end{minipage}\hfill
\begin{minipage}[b]{0.33\linewidth}
\includegraphics[width=\textwidth]{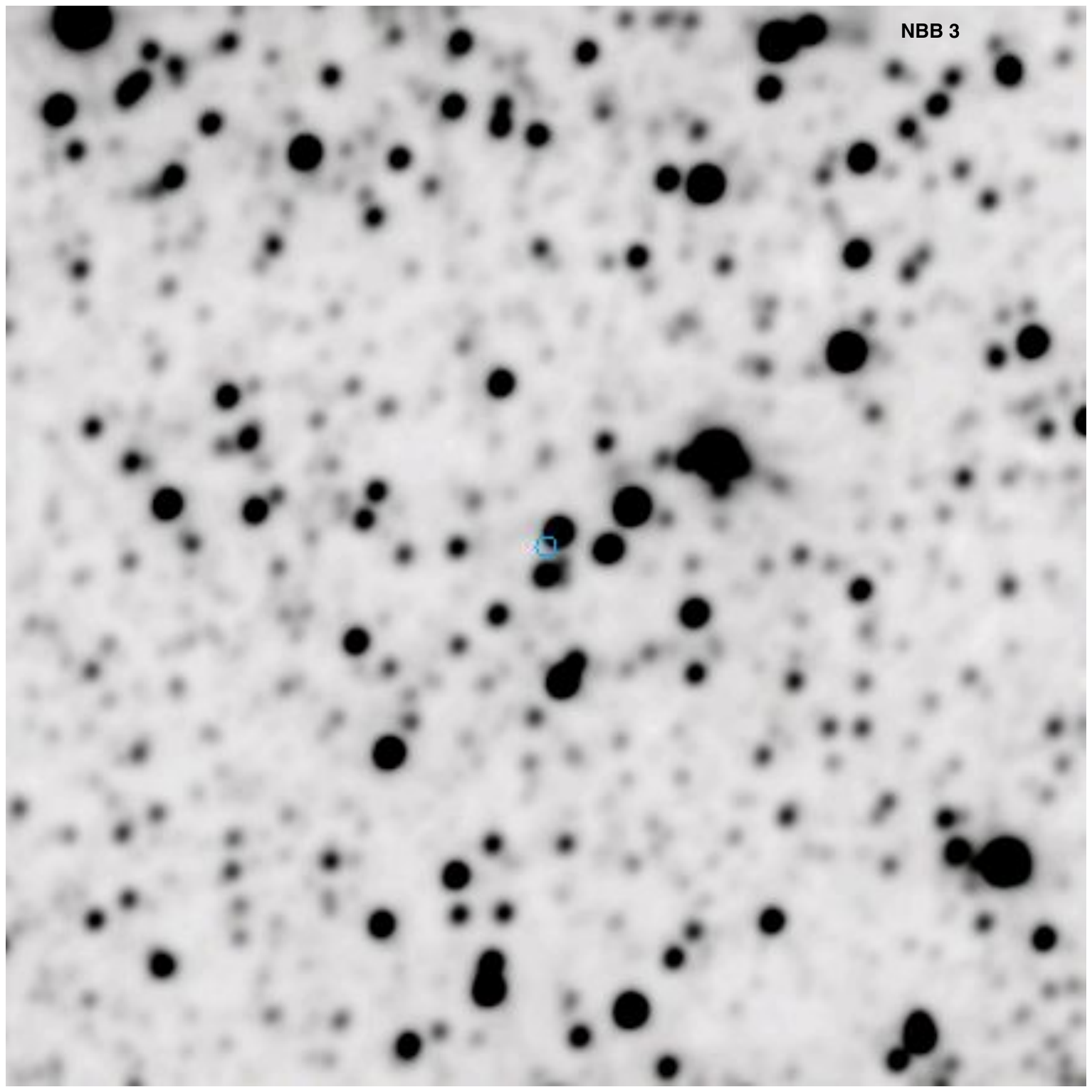}
\end{minipage}\hfill
\begin{minipage}[b]{0.33\linewidth}
\includegraphics[width=\textwidth]{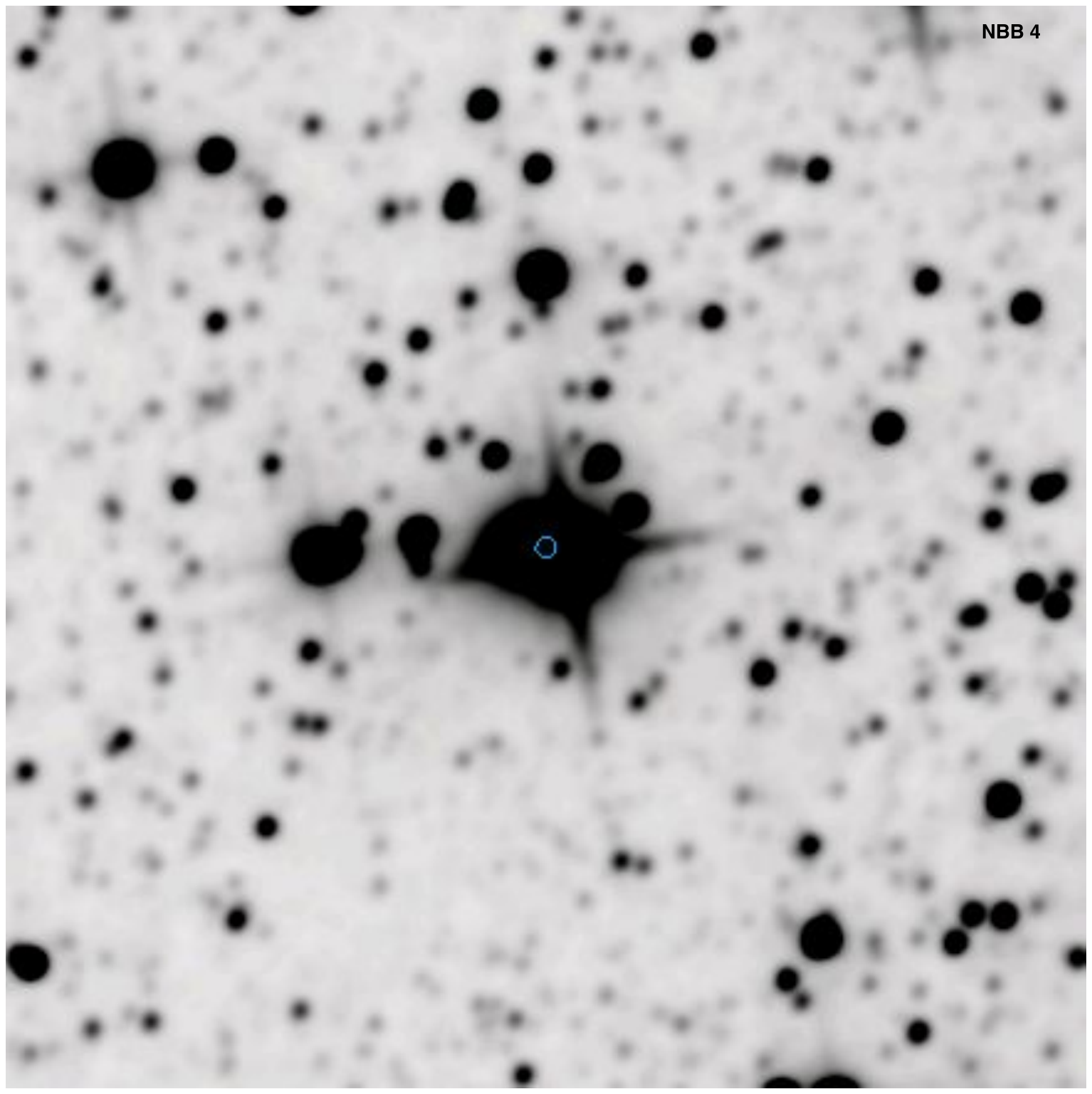}
\end{minipage}\hfill
\begin{minipage}[b]{0.33\linewidth}
\includegraphics[width=\textwidth]{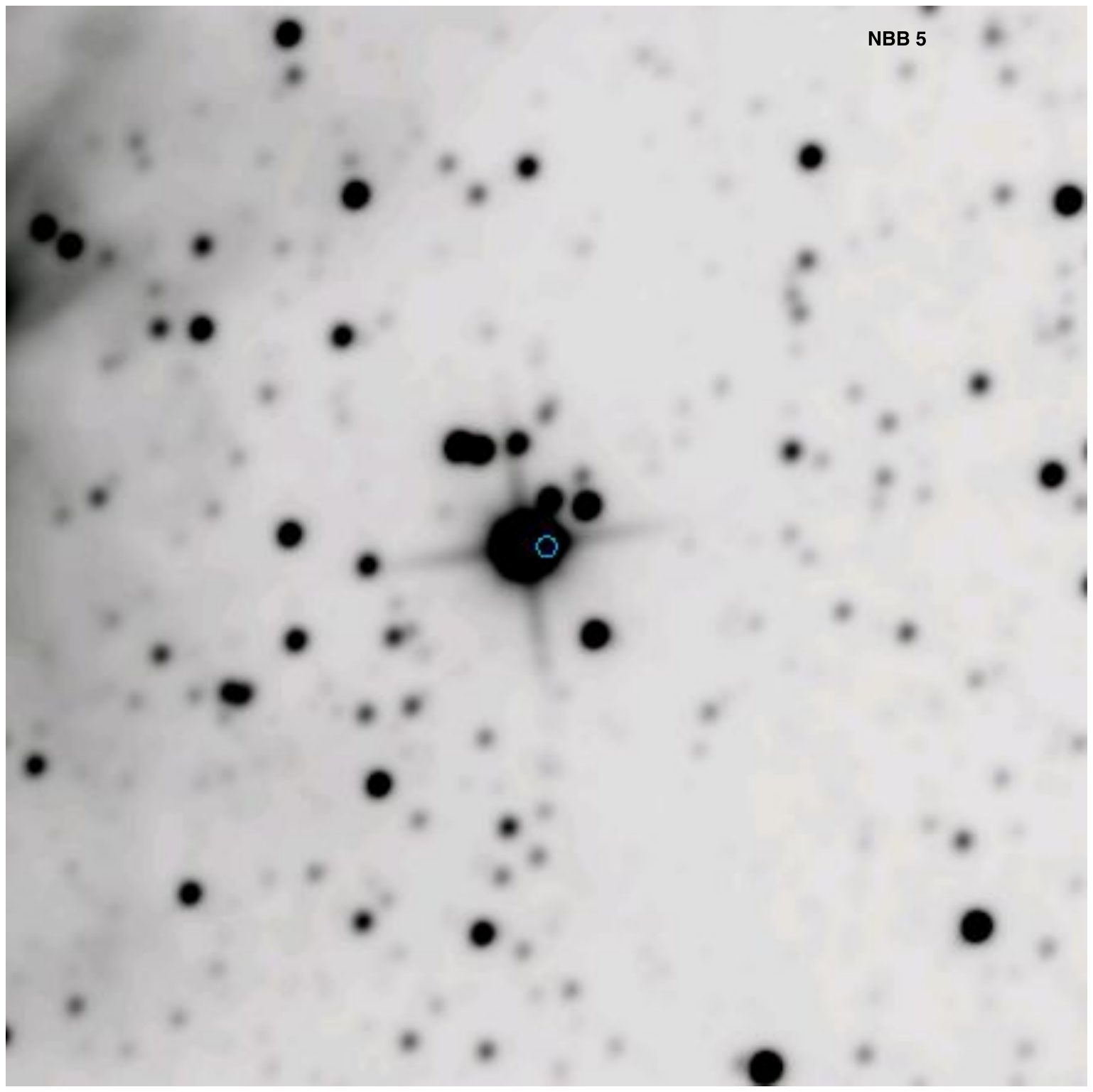}
\end{minipage}\hfill
\caption[]{WISE extractions in the W2 band ($10' \times 10'$) of the  five newly found embedded clusters. Upper panels  from left:  NBB 1, NBB 2 and NBB 3. Lower panels from  left: NBB 4 and NBB 5.}
\label{figXX}
\end{figure*}

\begin{figure*}
\begin{minipage}[b]{0.33\linewidth}
\includegraphics[width=\textwidth]{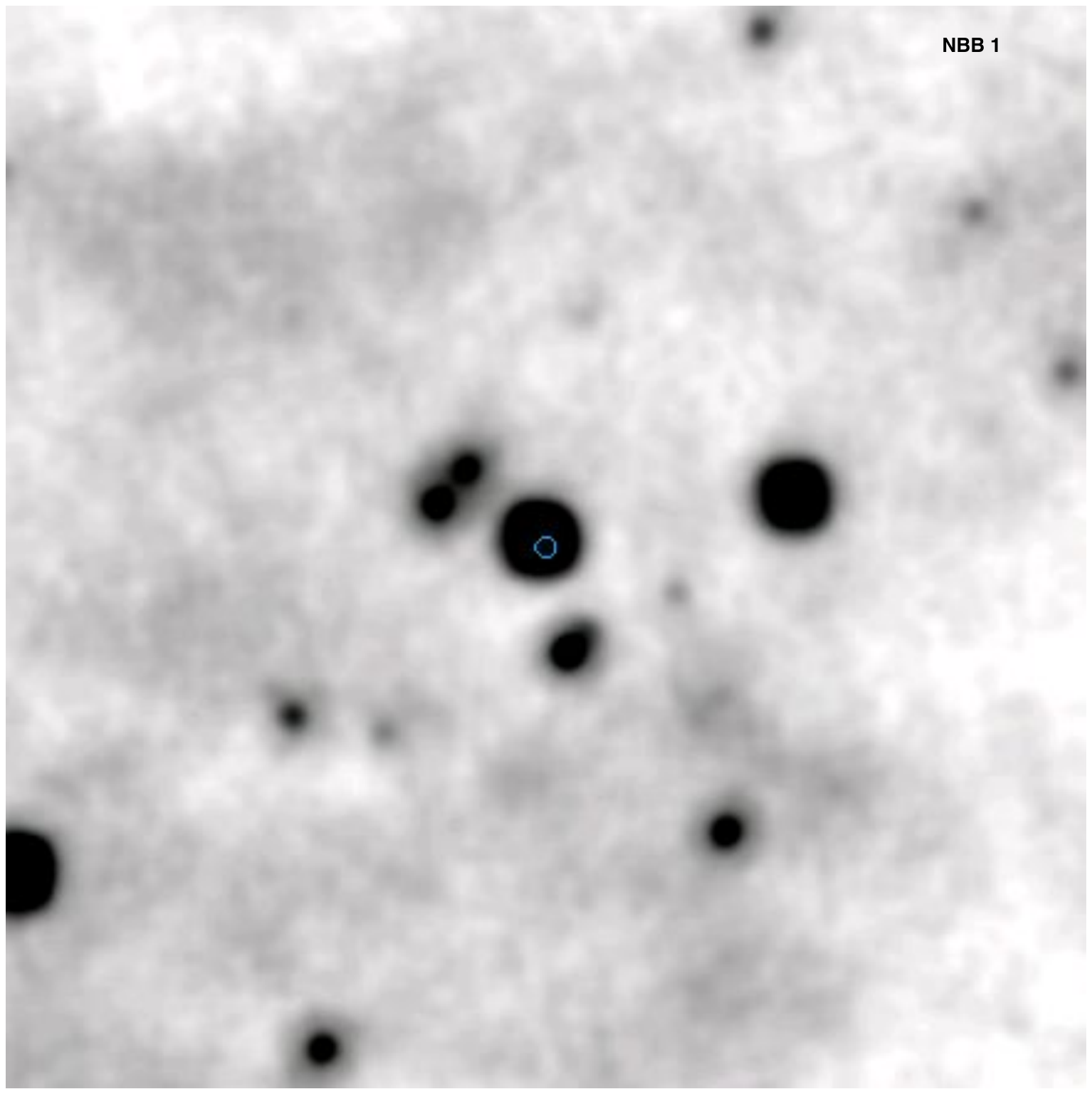}
\end{minipage}\hfill
\begin{minipage}[b]{0.33\linewidth}
\includegraphics[width=\textwidth]{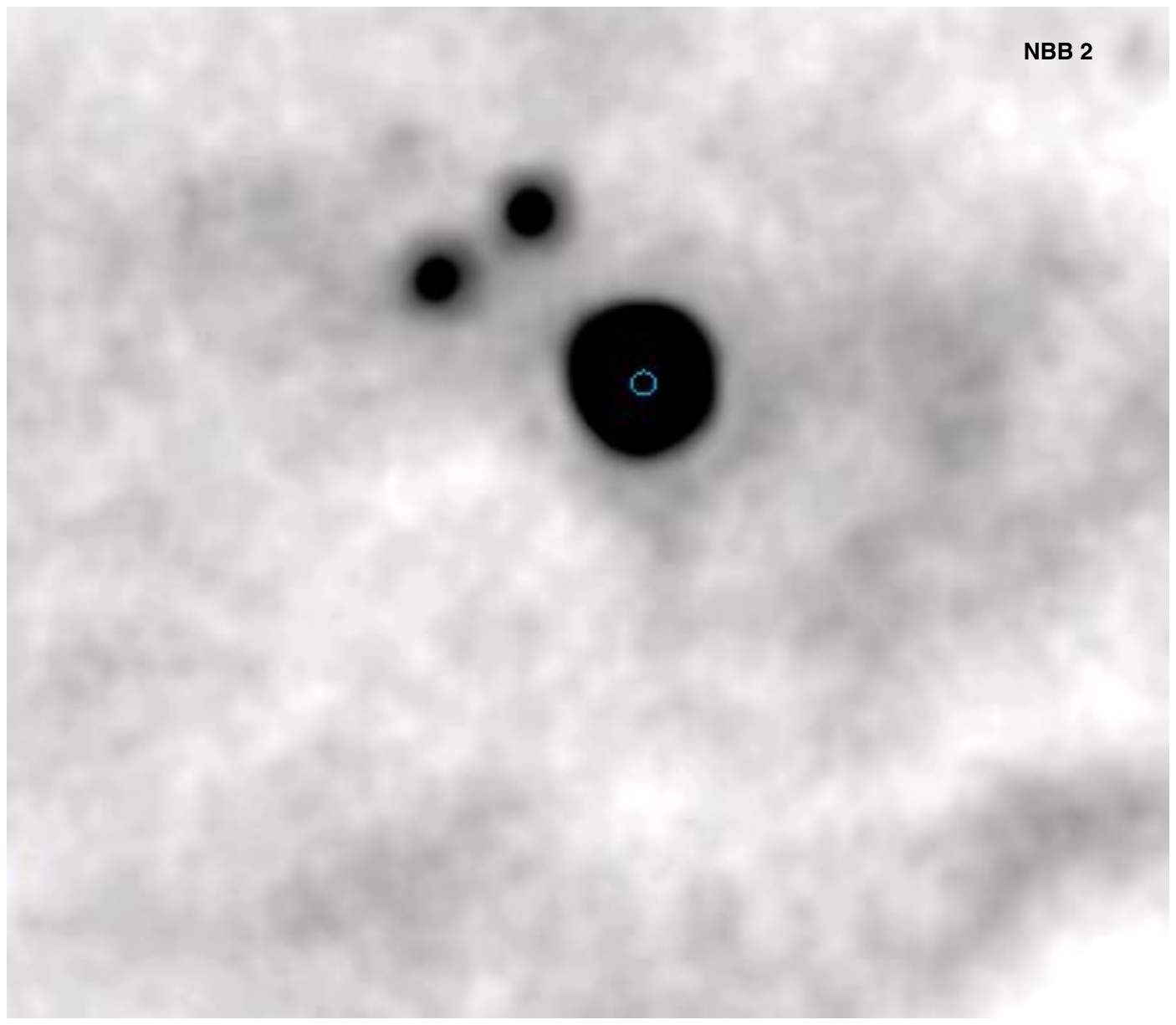}
\end{minipage}\hfill
\begin{minipage}[b]{0.33\linewidth}
\includegraphics[width=\textwidth]{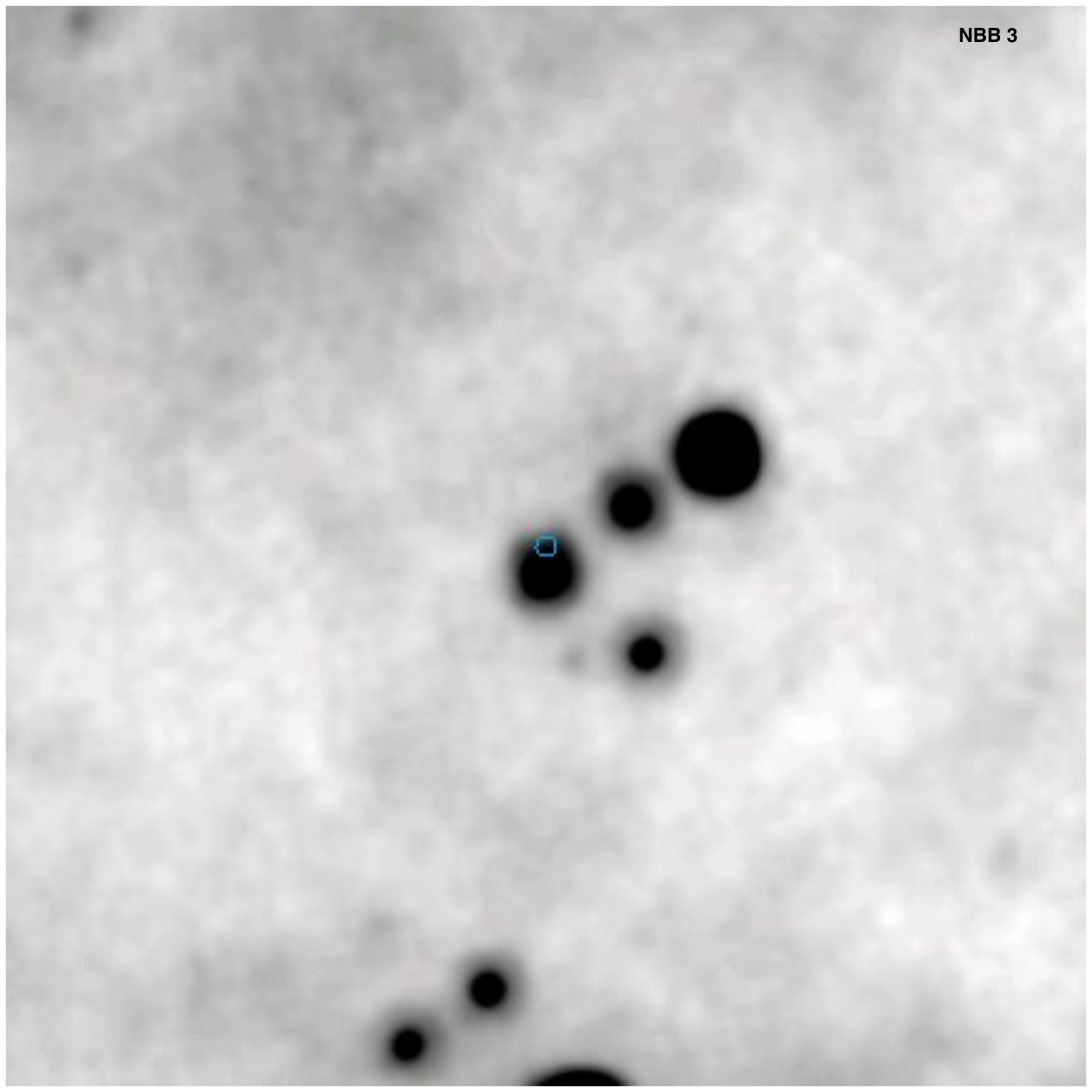}
\end{minipage}\hfill
\begin{minipage}[b]{0.33\linewidth}
\includegraphics[width=\textwidth]{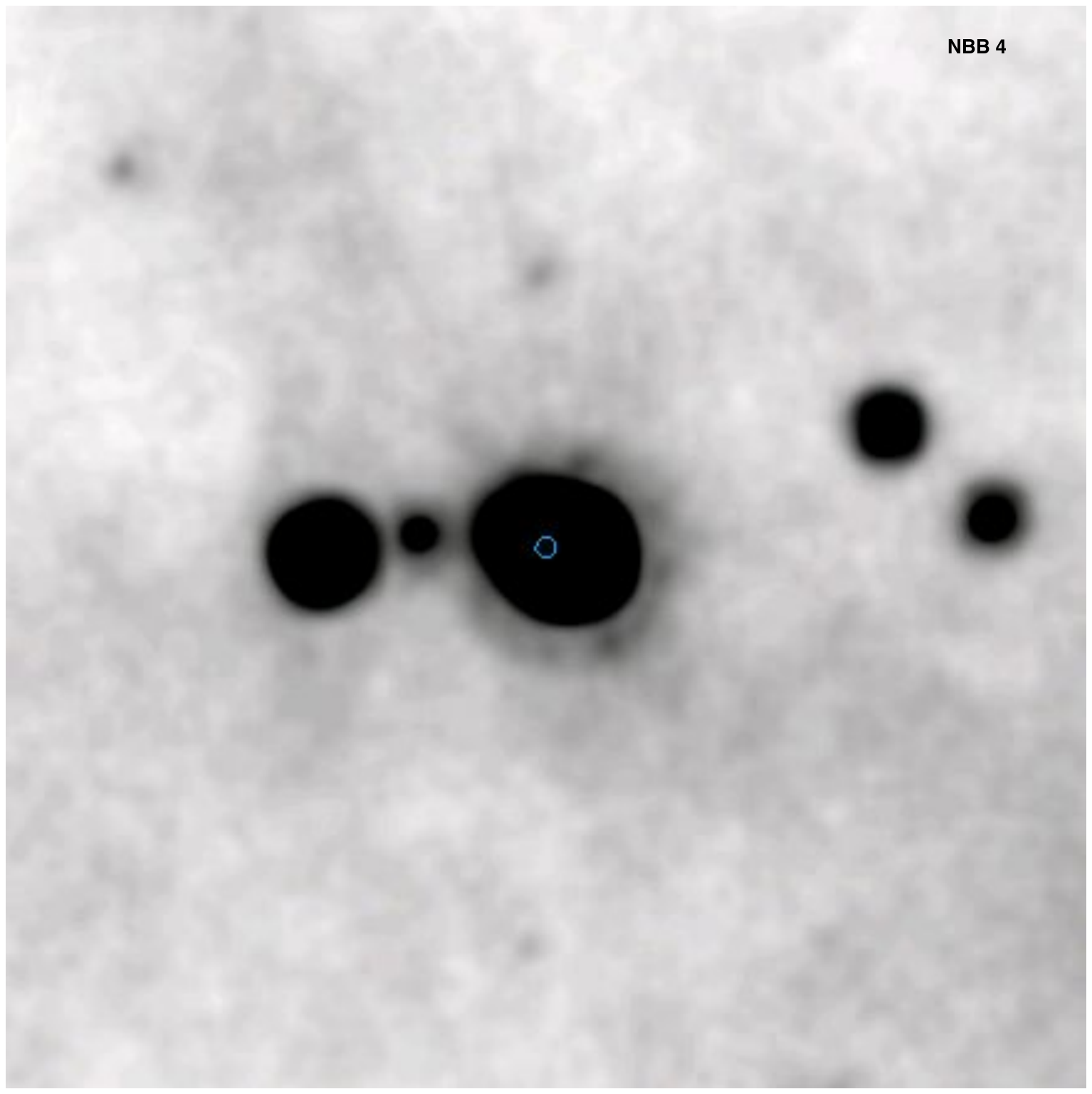}
\end{minipage}\hfill
\begin{minipage}[b]{0.33\linewidth}
\includegraphics[width=\textwidth]{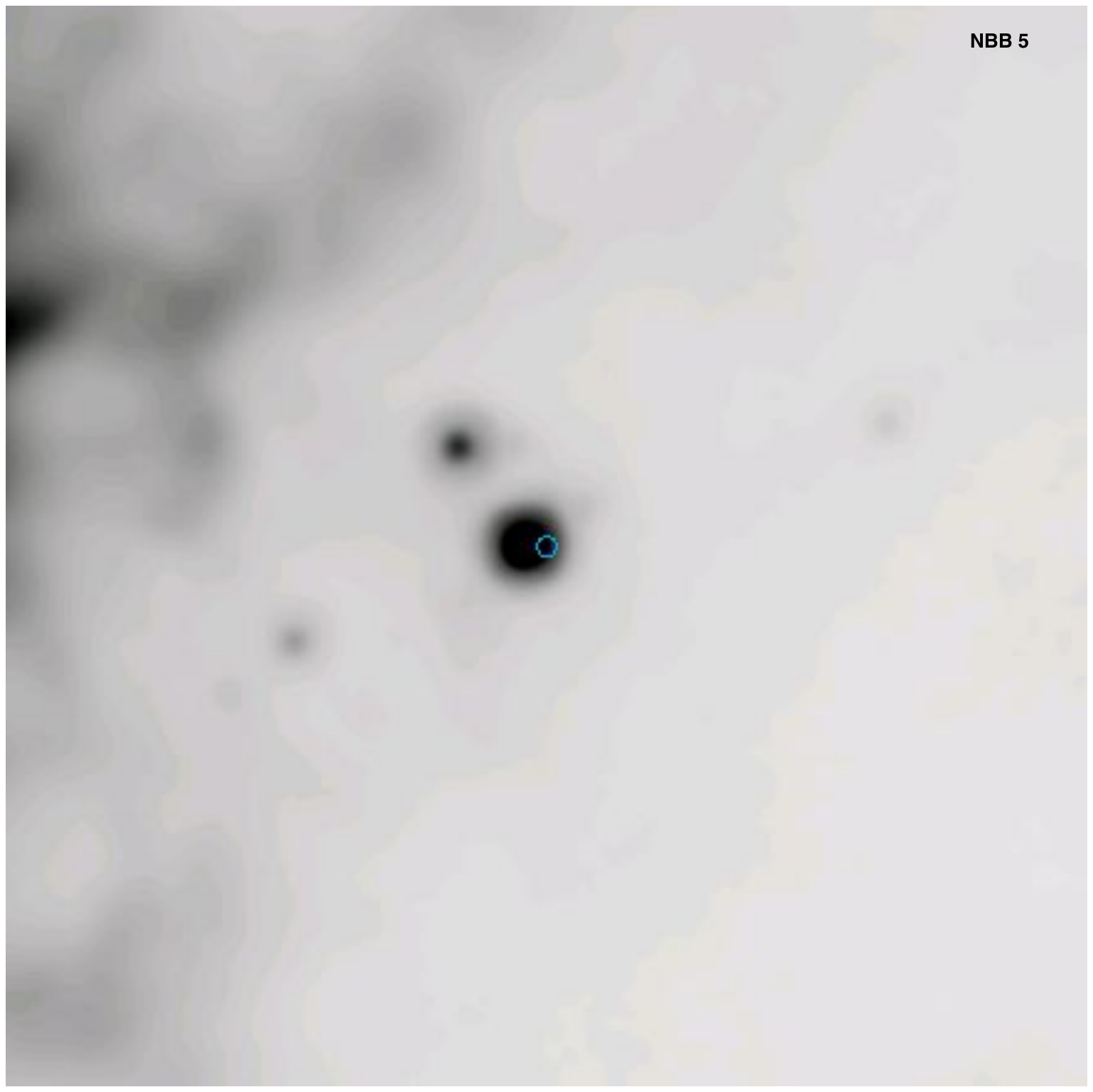}
\end{minipage}\hfill
\caption[]{ W4 band  extraction ($10' \times 10'$) of the  five newly found embedded clusters showing dust effects. Upper panels  from left:  NBB 1, NBB 2 and NBB 3.  Lower panels from  left: NBB 4 and NBB 5.}
\label{figXXX}
\end{figure*}

IC\,5146 is a reflection and emission nebula in the Cygnus constellation centred at about $l= 94\,^{\circ} 40'$, $b = -5\,^{\circ} 50'$. To the east of the cloud there occurs a  Streamer. The Streamer region corresponds to a long dark nebula discussed by Wolf (1904). Hubble (1922) classified it in the optical as continuous, whereas Minkowski (1947) listed the nebula to have H$\alpha$ emissions. Both are right, since IC\,5146 is a transition case between a reflection nebula and an HII region. Many surveys at different wavelengths have been carried out in this region and associated clouds (e.g. Kramer et al., 2003).
Fig. \ref{IC5146} shows an optical image of the IC\,5146 nebular and the Streamer ($\sim 1.7\,^{\circ} \times 0.8\,^{\circ}$) extending to the east. In the present study, one of us (E.B.) discovered 5 new ECs in the area of IC\,5146 using WISE (Fig. \ref{figXX} and \ref{figXXX}). The W2 images in Fig. \ref{figXX} enhance the stellar and protostellar components of the clusters, while  a W4  is sensitive to extended dust emission. Sources brighter in W4 than in W2 are certainly due to the presence of dust envelopes or disks. NBB\,5 is projected close to IC\,5146 and all the clusters candidates apperar to be immersed in cold molecular material. The central coordinates of these ECs are given in Table \ref{tab1}, together with the estimated angular radii. Although the cluster stellar concentrations may not be evident in Figs. \ref{figXX} and \ref{figXXX}, the embedded clusters, in general,  have conspicuous Radial Density Distributions in Fig. \ref{rdps}.

The IC\,5146 region has an important population of PMS stars. Embedded sources, sometimes known as protostars are optically invisible young stars. However, they are observed in the infrared  and the majority is associated with the cloud (see Sect. 3).We will explore IC\,5146 and the 5 ECs in the nebular complex using Colour-colour diagrams and CMDs (Bonatto \& Bica, 2007; Camargo, Bica \& Bonatto, 2015). In this work, IC\,5146 refers either to the nebula or the cluster.

\begin{table}
\caption[]{The embedded cluster IC 5146 and the 5 newly found ECs along the Streamer}
\label{tab1}
\renewcommand{\tabcolsep}{1.3mm}
\begin{tabular}{lccccccc}
\hline\hline
Name & $l$ & $b$ & $\alpha$ & $\delta$ & Radius & Number of stars \\
        & (deg)  & (deg) & (h:m:s) & ($^{o}:  '  : ''$)  & ($'$) &  (stars) & \\
\hline  
IC 5146 Cl       & $94.38$ & $-5.49$ & $21:53:26$ &$47:16:11$  & $7$ & $162$\\
NBB\,1$^\dagger$ & $93.49$ & $-4.72$ & $21:46:28$ &$47:18:06$  & $2$ & $47$ \\
NBB\,2$^\dagger$ & $93.53$ & $-4.26$ & $21:44:52$ &$47:40:32$  & $5$ & $24$ \\
NBB\,3$^\dagger$ & $93.60$ & $-4.10$ & $21:44:54$ &$47:46:00$  & $5$  & $31$\\
NBB\,4$^\dagger$ & $93.76$ & $-4.64$ & $21:47:22$ &$47:32:10$  & $8$ & $45$ \\
NBB\,5$^{\dagger\star}$ & $94.24$ & $-5.43$ & $21:52:33$ &$47:13:44$  & $3$ & $23$ \\
\hline
\end{tabular}
\begin{itemize}
\item $^\star$ close to IC\,5146 
\end{itemize}
\end{table}

%melhorar a figura usando WISE atlas; mudar a escala talvez

\section{The Data and Tools}
\label{sample}

The Infrared Array Camera (IRAC) was built at the NASA Goddard Space Flight Center (Fazio et al. 2004). IRAC is the mid-infrared camera on the Spitzer Space Telescope (Werner et al. 2004), with four arrays, or channels, simultaneously taking data in two separate fields of view. The four channels are referred to in this paper with their standard labels of $3.6, 4.5, 5.8$ and $8.0\mu m$ for channels $1, 2, 3, 4$, respectively, as described in the IRAC documentation and by Fazio et al.(2004). It is a powerful survey instrument because of its sensitivity, large FOV and four-colour imaging.
IRAC on Spitzer (e.g. Werner et al. 2004 and Fazio et al. 2004) has the potential to extend our understanding of disk evolution and star formation by detecting optically obscured, deeply embedded young stars and protostars. The emission is detected from their disks, and, at earlier stages, from their infalling envelopes (Lada \& Lada, 2003). The great advantage of IRAC over ground-based telescopes is its sensitivity in the $3-8 \mu m$ bands that contain relatively little contribution from stellar photospheres as compared to disks and envelopes. It is important to understand this colour space and use it to identify young stars of various evolutionary classes (Allen et al. 2004). %\citet{Allen et al.,2004}.
As an embedded cluster IC\,5146 still remains partly embedded in the gas and dust. The present  discovery of $5$ ECs in the area (Sect.2) further demands more detailed studies around IC\,5146.  We work with the 2MASS ($J, H, K_{s}$) and Spitzer (Ir1, Ir2, Ir3 and Ir4) photometry, which provide the spatial and photometric uniformity required for wide extractions. We employ the 2MASS ($J, H, K_{s}$) and Spitzer (Ir1, Ir2, Ir3 and Ir4) photometries, which provide considerably higher resolution than WISE. The sources in the 2MASS data are also cross-identified with a $0.5''$ search radius. Only sources detected in all four bands (W1, W2, W3 $\sigma \leq 0.1$ and W4 $\sigma \leq 0.5$) were initially considered. However, we realized that the W4 sources with profile magnitude  $> 8$ should be taked with caution (Jarrett et al. 2011 and Wright et al. 2010). We therefore limited our search for infrared excess to the range $3-12\mu m$. 
We did not use the WISE catalog flags to further filter our candidate list because we found them in general unreliable for our purposes. In particular, sources were often not flagged as extended or questionable, while no point source actually existed. The source classification with WISE data indicated source confusion in the central region and saturation effects, preventing identification of YSO candidates.  We conclude that WISE may not be suitable to classify YSOs in regions more distant than  the IC\,5146 complex. In principle, a random list of coordinates that falls inside a relatively smooth region of nebulosity might be $100\%$ classified as reliable.  Counterparts in Spitzer would support that. No attempt was made to compare  measured flux levels among WISE, 2MASS and Spitzer detections. This would require a surface brightness measurement that WISE is not calibrated to produce. As a consequence, we adopted 2MASS as standard catalog because the isochrones are well determinated (Bressan et al.2012).

\begin{figure}
\resizebox{\hsize}{!}{\includegraphics{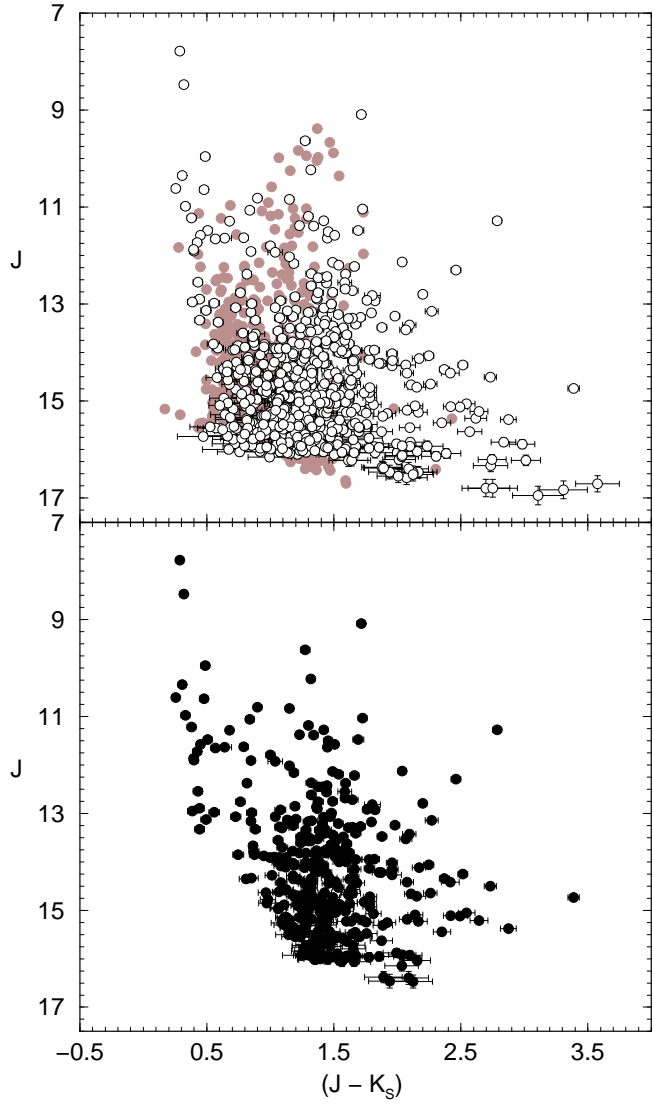}}
\caption{Decontamination example for IC5146 Cluster. Top panel: the observed photometry extracted for R=7' (open circles) is superimposed on the same area sky (grey). Bottom: decontaminated CMD.}
\label{decontaminated}
\end{figure}

Within this perspective, our group has been developing analytical tools for 2MASS photometry (Skrutskie et al., 1997) that allow us to statistically disentangle cluster evolutionary sequences from field stars in CMDs for IC\,5146 and the Streamer (Bonatto \& Bica, 2007). The decontamination procedure is illustrated in Fig. \ref{decontaminated} for IC\,5146. Decontaminated CMDs like in Figs. \ref{cmds} for IC\,5146 were used to investigate the nature of star cluster candidates and derive their astrophysical parameters (Bonatto \& Bica, 2006; Bonatto et al. 2008). We apply the statistical algorithm described in Bonatto \& Bica (2007b) to quantify the field-star contamination in the CMDs. The algorithm uses relative star-count desity together and colour/magnitude similarity between cluster and comparison field are taken  simultaneously. It measures the relative number densities of probable field and cluster stars in cubic CMD cells whose axes correspond to the J magnitude and the $(J-H)$ and $(J-K_{s})$ colours. The algorithm: (i) divides the full range of magnitude and colours covered by the CMD into a $3D$ grid, (ii) calculates the expected number density of field stars in each cell based on the number of comparison field stars with similar magnitude and colours as those in the cell, and (iii) subtracts the expected number of field stars from each cell. The algorithm is responsive to local variations of field-star contamination (Bonatto \& Bica 2007b). The adopted cell dimensions are large enough to allow sufficient star-count statistics in the cells and small enough to preserve the morphology of the CMD evolutionary sequences. For a representative background star-count statistics we use the ring located within $R_{inf} \leq R \leq R_{ext}$ around the cluster centre as the comparison field,where $R_{inf}$ usually represents twice the RDP radius. As extensively discussed in Bonatto \& Bica (2007b), differential reddening between cluster and field stars might be critical for the decontamination algorithm. Large gradients would require large cell sizes or, in extreme cases, preclude application of the algorithm. Basically, it would be required, cell size $\delta J = 1.0$ and $\delta(J-H) = \delta(J-K_{s}) = 0.2$ between cluster and comparison field for the differential reddening to affect the subtraction in a given cell (e.g. Bonatto \& Bica, 2008b). As a rule, a star cluster structure is studied by means of the stellar radial density profile (RDP). Usually, star clusters have RDPs following a power law profile (King, 1962; Wilson, 1975; Elson et al., 1987). However, ECs are often heavily embedded in their embryonic molecular cloud, which may absorb the near background and part of the cluster member stars. As a result, some ECs may present RDPs with decreasing density towards the cluster centre or bumps and dips (Lada \& Lada, 2003; Camargo, Bonatto \& Bica, 2011, 2012; Camargo, Bica \& Bonatto, 2013). Some RDP irregularities appear to be intrinsic to the embedded evolutionary stage and related to a fractal- like structure (Lada \& Lada, 2003; Camargo, Bonatto \& Bica, 2011, 2012; Camargo, Bica \& Bonatto, 2013).The Radial Density Profiles and CMDs (Figs. 5 and 6) determined in this study are consistent with previous work of Embedded Clusters (e.g. Bonatto \& Bica, 2009, Bonatto \& Bica, 2010).

Decontamination is a very important step in the identification and characterization of star clusters. Different approaches (Mercer et al.,2005) are based essentially on two different premises. The first relies on spatial variations of the star-count density, but does not take into account CMD evolutionary sequences. Alternatively, stars of an assumed cluster CMD are subtracted according to similarity of colour and magnitude with the stars of an equal-area comparison field CMD (as illustrated for the IC\,5146 in Fig. \ref{decontaminated}). In general, CMDs of ECs are dominated by PMS stars, so decontamination is important for avoiding confusion with the red dwarfs of the Galactic field and background reddened stars. The IC\,5146 cluster 2MASS Color-color diagram (Fig. 7) and that of Streamer (Fig. 10) are similar, which may suggest a similar absorption law due to similarity between the medium and the structure. A summary of different decontamination approaches is provided in Bonatto \& Bica (2009, 2010).

With this setup, the subtraction efficiency, i.e. the accumulated difference between the expected number of field stars (which may be fractional) and the number of stars effectively subtracted (integer), over all cells is higher than $90\%$ in all cases. The 2MASS decontamination results are given in Fig.6. In short, the cluster decontamination was carried out with the usual background circumnuclear ring geometry that best described it. (e.g. Bonatto \& Bica, 2011). These tools provided consistent RDP with observed profile of the sources and CMD results (Figs. 5 to 7).

%Finally, we show in Figs. \ref{fig_8} and \ref{fig_9} the matched (according to stars in common) samples of 2MASS and Spitzer. The shown CMDs are in Spitzer bands. The IC\,5146 cluster is considerably more reddened than the field. The CMDs suggest different cluster and field sequences (Fig. \ref{decontaminated}). The CMD morphology such as that of  the Streamer (Fig. \ref{fig_str}) shows similar features as the IC\,5146 PMS and possibly some Main-Sequence (MS) stars. The cluster Colour-colour diagram appear to follow the same absorption law as that of the field (Fig. \ref{ccs}).\\ 
%The 2MASS photometry was superimposed with the PADOVA of the PARSEC version ($1-5$ Myr) (Bressan et al., 2012). The MS stars are clearly described .The PMS appears to have a wider reddening range than the isochrone distribution. The panels corresponding to the Spitzer data show the MS counterparts. The derived parameters are in Table \ref{tab2} for IC\,5146 and the additional objects (Sect. ~\ref{CMDs} and ~\ref{disc}) . Table \ref{tab2}, in particular, compares distance determinations in the literature for IC\5146, together with the one found in the present work.

The MS stars of IC 5146 in the 2MASS CMDs  (Fig. \ref{cmds}) correspond to the blue isochrone, and they  appear to have  comparable reddening as the field blue  envelope of stars.
The redder PMS stars populate the redder parts of the CMD, as  expected. The matched 2MASS-Spitzer diagrams (Fig. 8) of IC 5146 indicate a mixture of these populations, and/or some
splitting between cluster and field sources. The 2MASS and Spitzer CMDs and colour-colour diagrams of the Streamer (Fig.10) show  similarities with thoseof IC 5146, with some MS stars and numerous PMS stars. The 2MASS photometry superimposed with the PADOVA of the PARSEC version $(1 - 5 Myr$) (Bressan et al. 2012), indicates that  the MS stars are clearly described,indicating an ageing cluster. The PMS shows  a wider colour range than the isochrone distribution in Fig. 6. This implies that part of the sources are  extremely absorbed or have excess emission owing to envelopes or disks.
The derived parameters from the superimposed isochrones are given in  Table 2 for IC 5146 and the additional objects (Sect. 4 and 5). Table 3 compares distance determinations in the literature for IC 5146 to that derived in the  present study. Our distance  agrees well with the larger values, e.g. in Herbig \& Dahm (2002).

%figura CMD avermelhamento

\section{The 5 newly found embedded clusters}
\label{CMDs}

We found 5 new ECs (Section \ref{history}) in the WISE images of the Streamer to the east of IC\,5146. WISE is a survey covering the entire sky in the infrared bands W1, W2, W3 and W4 centred at $3.4$, $4.6$, $12$ and $22\mu m $ respectively. It provides efficient detections of star forming clusters (Majaess, 2013; Camargo, Bica \& Bonatto, 2015). There were two basic strategies to identify new EC candidates. The first consists of targeted searches around IR and radio sources, or in areas of interest (e.g., the Galactic Bulge or known nearby star-forming regions; i.e, Dutra \& Bica, 2000). The second is unbiased: it is based on techniques that apply sets of objective rules to identify overdensites in flux or stellar-number density in the pixel data or in the point-source catalogs, respectively. Our search focuses on the stellar density and gas/dust distribution to identify star clusters, stellar groups, and candidates. Furthermore, color-color diagrams have been known to be powerful tools in identifying the ECs. WISE images of the W2 band (Fig. \ref{figXX}) show probable MS and PMS stars of the clusters.The W3 and W4 bands mainly probe the circumstellar/interstellar dust emission. The Streamer located to the northeast of IC\,5146 and it has a filamentary elongated structure, probably as a consequence of the gas dynamics. The region might potentially form protoclusters.
Fig. \ref{IC5146} shows that 4 ECs are seen along the Streamer, while NBB\,5 is projected close to IC\,5146. The analysis of these ECs was made using the 2MASS data (Figs. \ref{cmds} and \ref{ccs}). Regarding NBB\,5, it is the least populous among our candidates (Figs. 2 and 3). But, its CMD morphology (Fig. \ref{cmds}) and colour-colour diagram (Fig. \ref{ccs}) are consistent with those of the remaining ECs. Thus, we include it as a possible EC in the region, but noting that deeper photometry would be required for a conclusion on its nature.

Historically, different approaches have been used to extract astrophysical parameters from isochrone fits. The simplest ones are based on a direct comparison of a set of isochrones with the CMD morphology and Colour-colour diagrams (Figs. \ref{cmds} and \ref{ccs}, respectively), while the more detailed ones include photometric uncertainties, binaries and metallicity variations. However, because of the 2MASS photometric uncertainties for the lower sequences, a more detailed approach for isochrone fitting might lead to overinterpretation. Nevertheless, some effects such as degree of flattening of the envelope or presence of outflow may change the PMS colours as well, especially at low densities (Whitney et al.,2003). Despite the changes already introduced in PARSEC code (Bressan et al., 2012) we point out that the theoretical isochrones do not take into account the circumstellar disk/envelope dust emission for PMS stars or YSOs, which is essential for modeling the earliest evolutionary phases of pre-main sequence stars. So, the fitted astrophysical parameters as reported in this work should be taken with caution.

Fundamental parameters are derived using the field decontaminated CMD fitted with PMS PADOVA isochrones (Bressan et al. 2012). The derived parameters (Table \ref{tab2}) are the observed distance modulus $(m-M)_J$ and reddening $E(J-H)$, which converts to $E(B-V)$ and $A_v$ with the relations $A_J/A_V=0.276$, $A_H/A_V = 0.176$, $ A_{K_s}/A_V=0.118$, $A_J=2.76 \times E(J-H)$ and $E(J-H)= 0.33 \times E(B-V)$ (Dutra, Santiago \& Bica, 2002). We applied the direct comparison of isochrones with the decontaminated CMD morphology (Fig. \ref{cmds}). The fits are  made \textit{by eye}, taking the combined MS and PMS stellar distributions as constraints, allowing as well differential reddening and photometric uncertainties. Such fits are traditionally used in the analyses of open and globular CMDs (e.g. Bonatto et al. 2010 and Barbuy et al. 2007). Automatic fits (e.g. Naylor et al. 2006) are also an important tool, but they may become uncertain facing contaminated CMDs with atypical stars outside the main evolutionary sequences and/or effects of differential reddening. Isochrones of Bressan et al (2012) with ages in the range $1- 20$ Myr are used to characterise the PMS sequence. Given the poorly populated MS, acceptable fits to the decontaminated MS morphology are obtained with any isochrone with age in the rande $1- 10$ Myr. The PMS stars are basically contained within this range of isochrones as well, which implies a similar age range as the MS. Thus, we take the $5$ Myr isochrone as representative solution, with a $3$ Myr spread. In addition, we considered the envelope adjustment along the MS as PMS (Mayne et al. 2007), just prior to the onset of hydrogen burning when they enter a quasi-equilibrium state as hydrogen ignition starts. This delays their arrival onto the MS, allowing variations due to photometric uncertainties and differential reddening (Bonatto \& Bica, 2010). We assume a constant total to selected absorption ratio $R_V=3.1$ (Cardelli, Clayton \& Mathis, 1989). The derived parameters are given in Table \ref{tab2}.

\begin{table}
\caption[]{Reddening, age and distance for IC\,5146, the Streamer and the 5 protoclusters}
\label{tab2}
\renewcommand{\tabcolsep}{1.8mm}
\begin{tabular}{lcccccc}
\hline\hline
Protocluster & $E(J-Ks)$ & $(m-M)_{J}$ & $A_{v}$ & $Age$ & $distance$ \\
                & (mag)  &   (pc)      &(mag)    & (Myr) & (kpc)  \\
\hline
IC\,5146          &$0.42 \pm 0.05$ & $11.40 \pm 0.13$ & $2.43 \pm 0.29$ & $5 \pm 3$ & $1.3 \pm 0.1$ \\
NBB\,1$^\dagger$  &$0.42 \pm 0.05$ & $10.94 \pm 0.13$ & $2.44 \pm 0.29$ & $5 \pm 3$ & $1.1 \pm 0.1$  \\
NBB\,2$^\dagger$  &$0.50 \pm 0.05$ & $12.42 \pm 0.13$ & $2.91 \pm 0.29$ & $5 \pm 3$ & $2.1 \pm 0.2$  \\
NBB\,3$^\dagger$  &$0.45 \pm 0.05$ & $11.94 \pm 0.13$ & $2.62 \pm 0.29$ & $5 \pm 3$ & $1.7 \pm 0.1$  \\
NBB\,4$^\dagger$  &$0.40 \pm 0.05$ & $11.95 \pm 0.13$ & $2.33 \pm 0.29$ & $5 \pm 3$ & $1.8 \pm 0.1$ \\
NBB\,5$^\dagger$  &$0.34 \pm 0.05$ & $11.23 \pm 0.13$ & $1.98 \pm 0.29$ & $5 \pm 3$ & $1.4 \pm 0.1$ \\ 
Streamer          &$0.49 \pm 0.05$ & $11.27 \pm 0.13$ & $2.85 \pm 0.29$ & $5 \pm 3$ & $1.2 \pm 0.1$ \\  

\hline
\end{tabular}
\begin{list}{Table Notes.}
\item $\dagger$ ECs centered according to Table \ref{tab1}.
\end{list}
\end{table}

Measurements of distance to the Streamer do not exist in the literature. Usually the distance of this region is assumed to be the same as that derived optically for B stars in the Cocoon Nebula, the HII region at the far eastern of the IC\,5146 complex (Elias, 1978; Dobashi et al. 1992 and Lada, Alves \& Lada, 1999). However, based on the relative lack of foreground stars, Lada et al.(1994) estimated a distance of only $400$pc to the cloud, as compared to our estimated value of $1200$ pc, thus closer to IC\,5146.

\begin{table}
\caption[]{Distance determination to IC 5146}
\label{tab3}
\renewcommand{\tabcolsep}{8.5mm}
\begin{tabular}{ccc}
\hline\hline
    Reference  &    Distance  & Notes  \\
                      & (pc)  \\
\hline
Walker (1959)  &   $ 1000 $            &   [1]   \\
Crampton \& Fisher(1974)  &   $960$   &   [2]  \\
Elias (1978)  &   $ 900 \pm 100$       &  [3]   \\
Forte \& Orsatti (1984)  &   $1000$   &  [4]  \\
Lada et al.,(1999)    &  $460 \pm 40$  &  [5] \\
Herbig \& Dahm (2002) &   $1400 \pm 180$  & [6]    \\
Herbig \& Dahm (2002) &   $1100 \pm 180$  & [7]  \\
Harvey et al.,(2008)  &   $950 \pm 80 $   & [8]  \\
This work  &  $1300 \pm 100$  & [9]  \\
\hline
\end{tabular}
\begin{list}{Table Notes - The adopted methods and references are:}
\item [1] Johnson \& Hilter (1956)-ZAMS; [2] Walborn (1972)-ZAMS; [3]Blaauw (1963)-ZAMS; [4] Balona \& Feast (1975)-ZAMS; [5] Star counts; [6] Allen et al.(1982)-ZAMS; [7] Jaschek \& Gomez (1998)-MS; [8] Harvey et al.(2008)-ONC ZAMS; [9] PMS and MS.
\end{list}
\end{table}

\section{Discussions}
\label{disc}

The majority of the embedded clusters do not survive the first few ten Myr, especially the low-mass ones, given the interplay among environmental conditions, star-formation efficiency, initial stellar dynamical state and mass fraction converted into stars or expelled. It is important to investigate the structural and photometric properties in this early phase. We report the discovery of $5$ ECs along the Streamer to the east of IC\,5146 and compare their CMDs and Colour-colour diagrams, suggesting a similarity of properties among them. 

The features in the Colour-colour diagrams of the Streamer and the 5 ECs are compatible with an age of $\sim$ $5$ Myr (Fig. \ref{ccs} and Table \ref{tab2}). In principle, the detailed distribution of stars within the PMS area in the CMD plane reflects the star formation history of the embedded population. The relative mean ages of young clusters can be established to greater precision by using a single or a consistent set of PMS models to extract the cluster ages (e.g., Haisch et al., 2001a and Bressan et al., 2012)

%Comparison with the isochrones derived from a single set of PMS tracks gives a cluster mean age of $ \sim 2$ Myr and an age spread of approximately 5 Myrs. 

\begin{figure}[h]
\resizebox{\hsize}{!}{\includegraphics{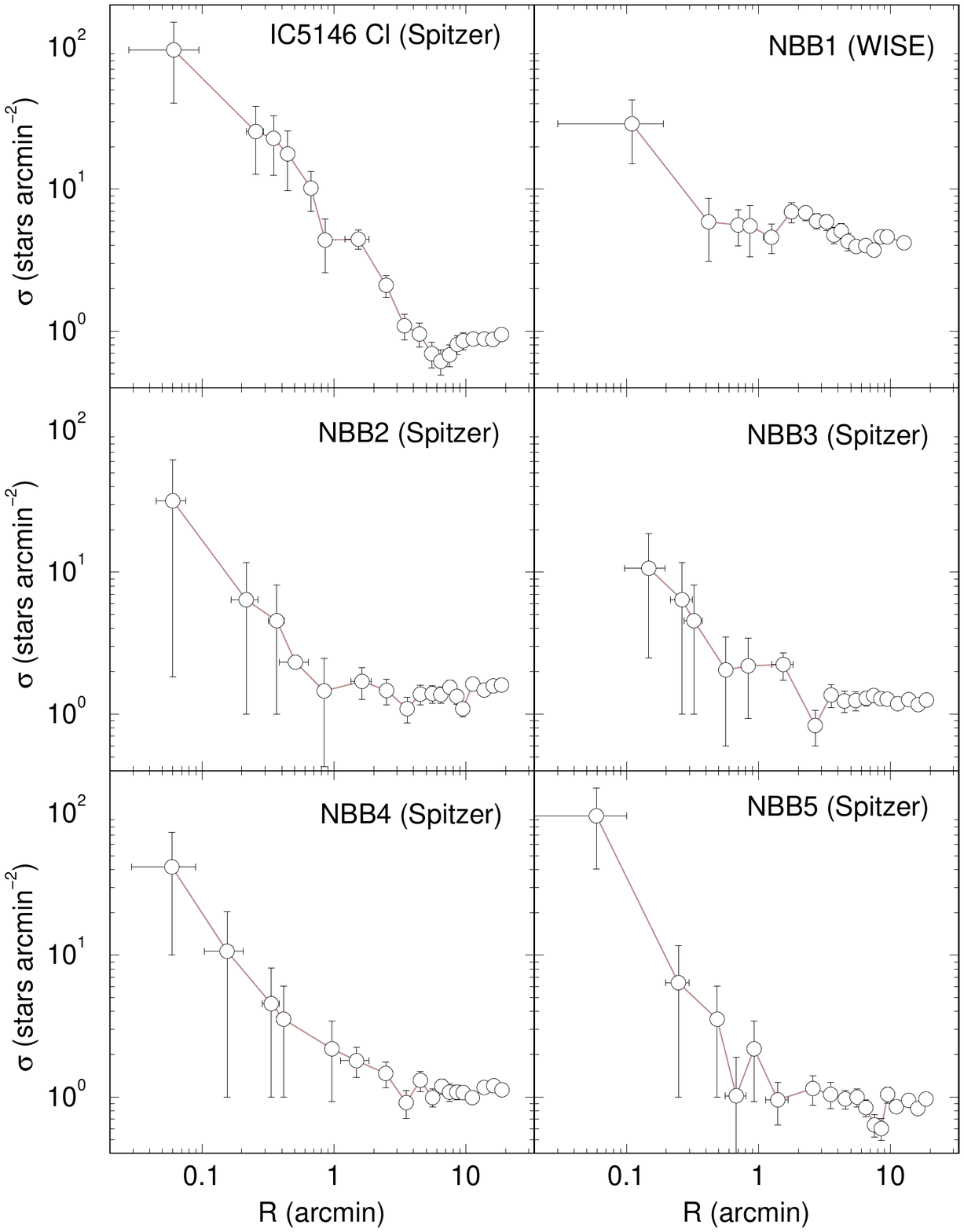}}
\caption{The radial density profiles of IC\,5146 and the newly found ECs.}
\label{rdps}
\end{figure}

\begin{figure}[h]
\resizebox{\hsize}{!}{\includegraphics{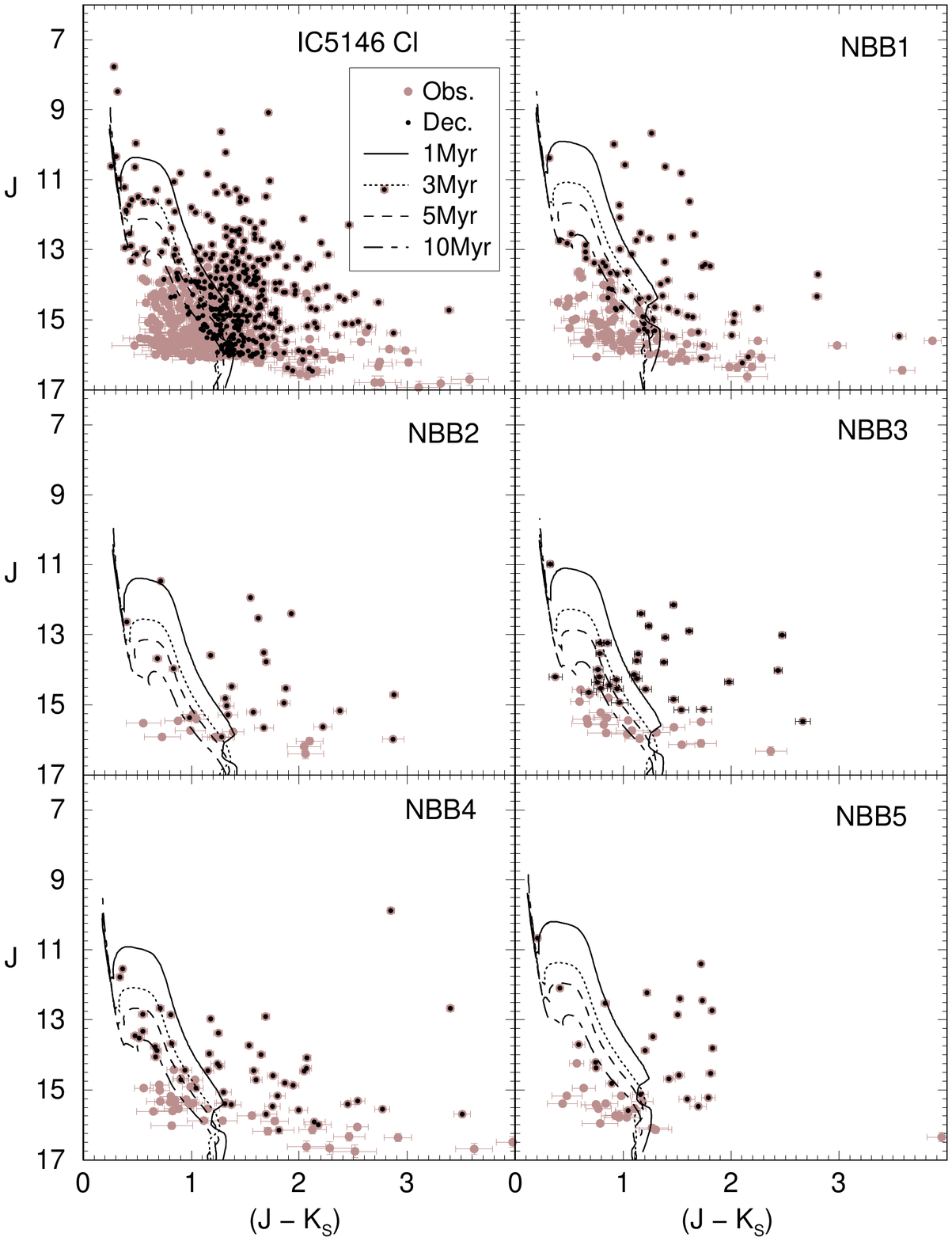}}
\caption{Decontaminated CMDs of the five newly found ECs and IC\,5146  fitted with MS and PMS PADOVA isochrones. Brown points correspond to the field stars. Black points are statistically subtracted cluster stars.}
\label{cmds}
\end{figure}

In all five new ECs (Table \ref{tab2}), the young stars identified by their excess emission in the mid-IR are possibly distributed in a range of distances. In contrast, the diameters of clusters by star counts are typically 4 pc or less (Lada \& Lada, 2003). This confirms that a significant fraction of gas in each star-forming region forms outside and surrounding the clustered region. The distances and uncertainties suggest that the NBBs clusters are associated physically. Possibly NBB\,2 is a bit far.

\begin{figure}[h]
\resizebox{\hsize}{!}{\includegraphics{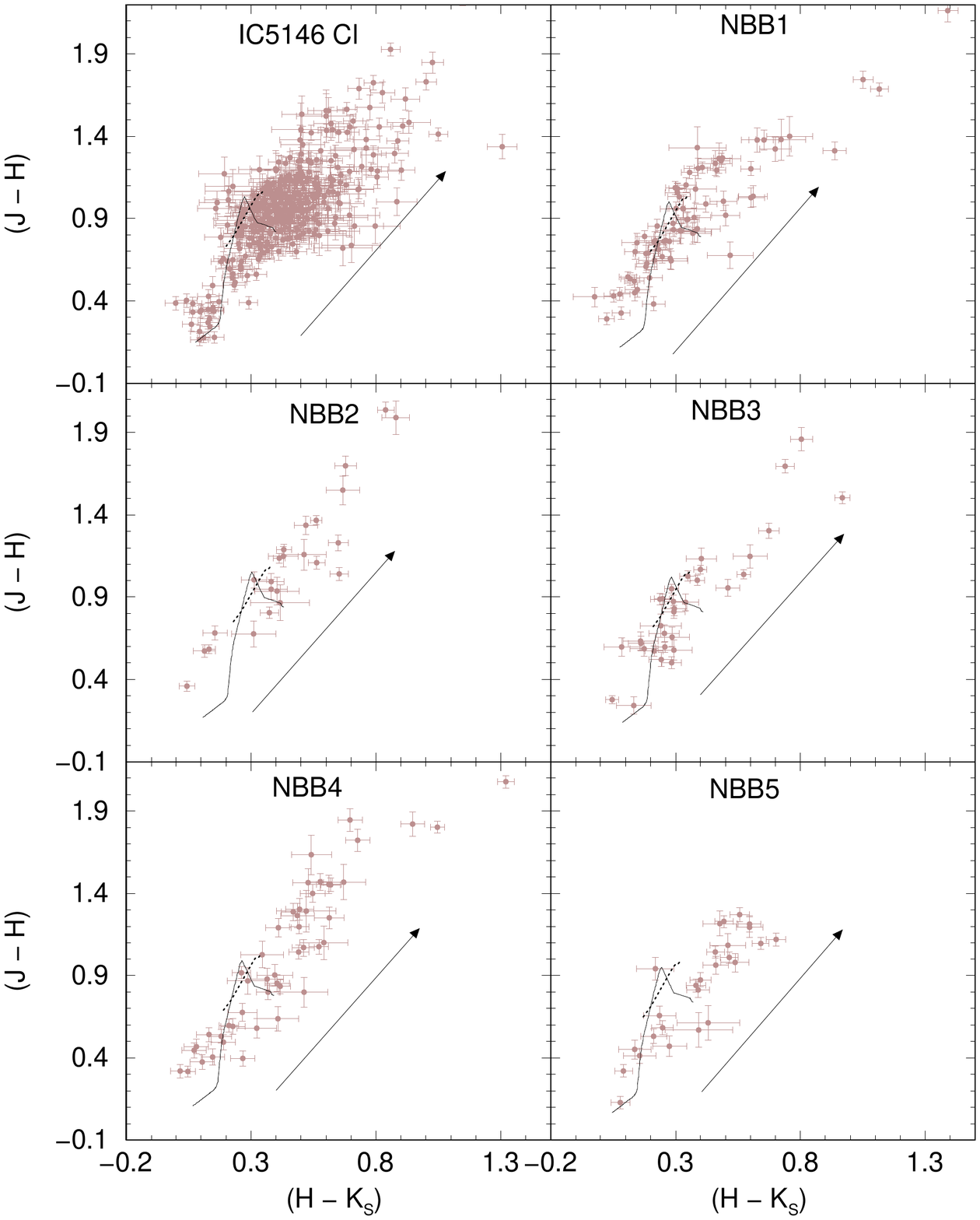}}
\caption{Decontaminated Colour-Colour Diagrams fitted with PADOVA isochrones for the IC\,5146 Cluster and the 5 newly found ECs. $A_{V}= 10$ mag.}
\label{ccs}
\end{figure}

The 2MASS CMDs (Fig.\ref{cmds}) and estimated number of stars in column 7 in Table \ref{tab1} provide that IC\,5146 is by far, the most populous in the sample.

Futhermore, we suggest that star formation in IC\,5146 is continuing in the halo, while the molecular gas has been dispersed toward the cluster core and Streamer. 

\begin{landscape}
\begin{figure}
\resizebox{\hsize}{!}{\includegraphics{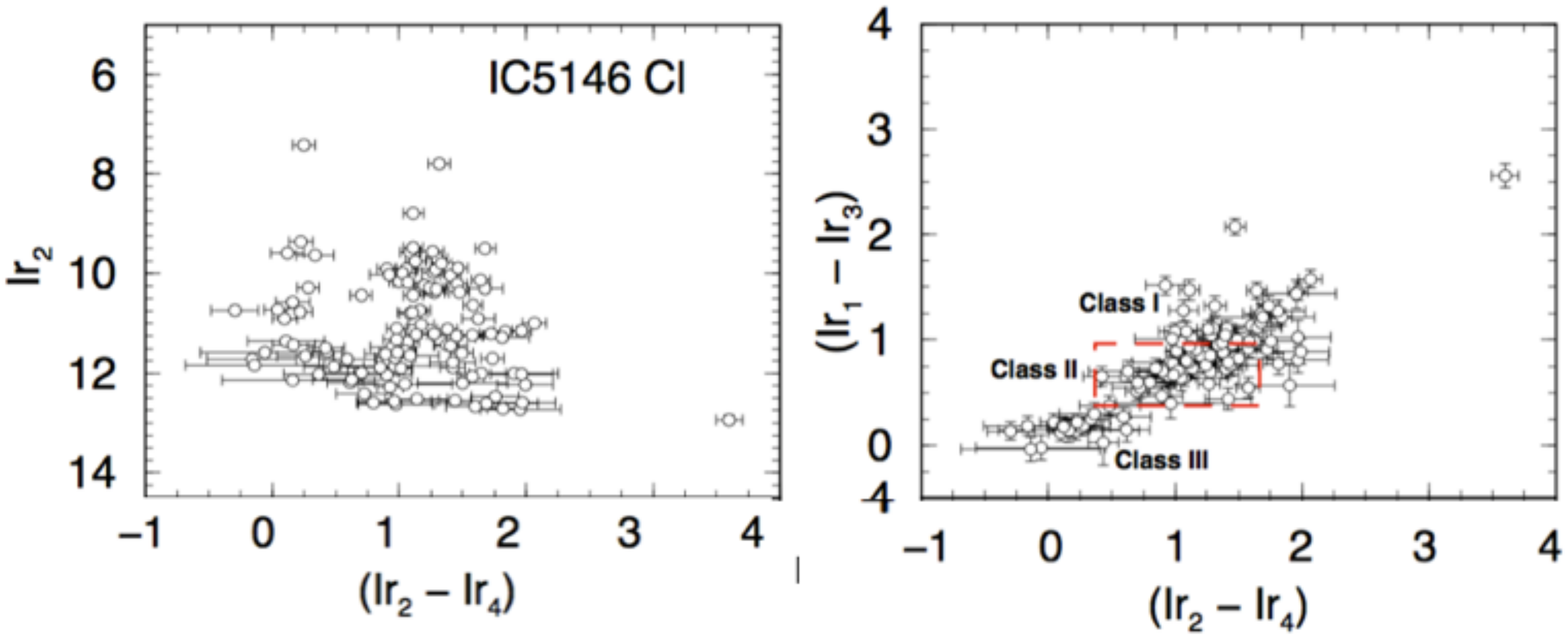}}
\caption{IC\,5146:  $Ir2\times(Ir2-Ir4)$ and $(Ir1-Ir3)\times (Ir2-Ir4)$. Points are matched stars with 2MASS and classification of the YSOs.}
\label{fig_8}
\end{figure}
\end{landscape}

\begin{figure}[h]
\resizebox{\hsize}{!}{\includegraphics{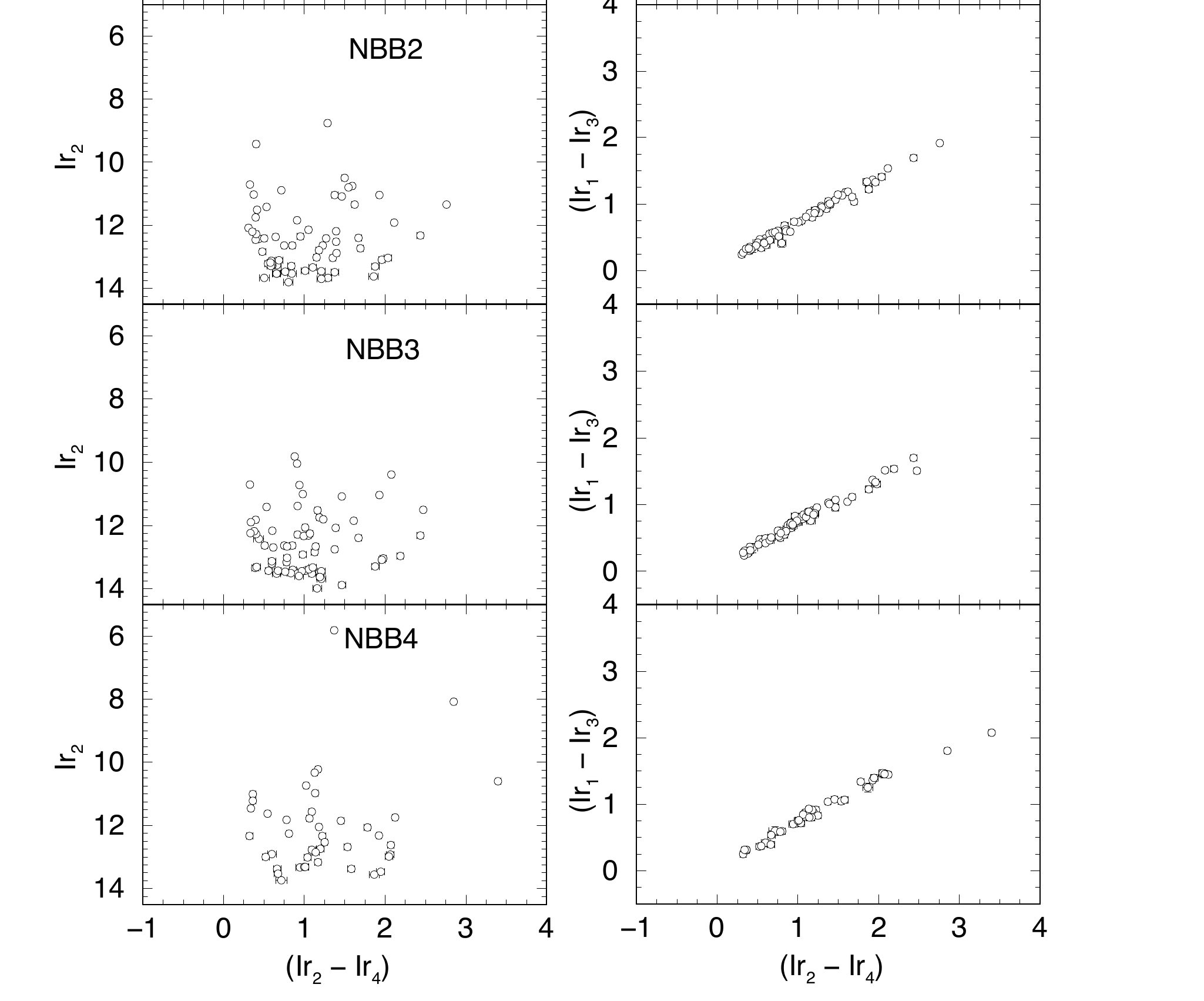}}
\caption{The 3 ECs : $Ir2\times(Ir2-Ir4)$ and $(Ir1-Ir3)\times (Ir2-Ir4)$. Points are statistically matched stars with 2MASS.}
\label{fig_9}
\end{figure}

Megeath et al. 2004 \& Gutermuth et al. 2004 have discussed YSO classifications. Fig. \ref{fig_8} shows matched IRAC  colors of IC\,5146, together with model loci by Allen et al. 2004. The data seem to cluster into three main regions: (i) a clump around  $(Ir2-Ir4) \times (Ir1-Ir3) = [0, 0]$ that contains mostly background/foreground stars and Class III sources with no intrinsic IR excess, (ii) a clump that occupies the Class II region (within the large red box), and a group that runs along the Class I locus. Sources that lie between the $(Ir2-Ir4) \times (Ir1-Ir3)$ colors of stellar photospheres and Class III stars $[0,0]$ and Class II sources $[0.4, 1.0]$ can be understood if we suggests the effects of dust grain growth and settling. In addition, there are a few sources that do not lie inside the Class II locus but do lie along the extinction vector and could therefore be reddened Class II objects. The most extreme points might be trace filaments and high opacity (dark even at $8 \mu m$), consistent with the large $A_{v}$ implied by this scenario. Alternatively, they could be extremely low luminosity Class I sources. Thus, the Class I stage (Fig. \ref{cmds} to \ref{fig_9}) is a later stage of protostellar collapse. 

The Class II stage (Fig. \ref{fig_8}) is characterized by the presence of excess of IR emission above that expected for a stellar photosphere. These stars can be inside it or lie along the extinction vector and could therefore be reddened. This is the signature of a classical T Tauri surrounded by an accretion disk.  Infall from the cloud has ceased owing to dispersal of the remnant infall envelope by the combined effects of infall and outflow (Allen et al. 2004; Megeath et al. 2004; Sung, Bessell \& Lee, 1997). The Class III stage can last for about $10^{7}$yr ending in a zero-age main-sequence star (Palla, 1999).

\begin{figure}[!htbp]
\centering
\resizebox{\hsize}{!}{\includegraphics{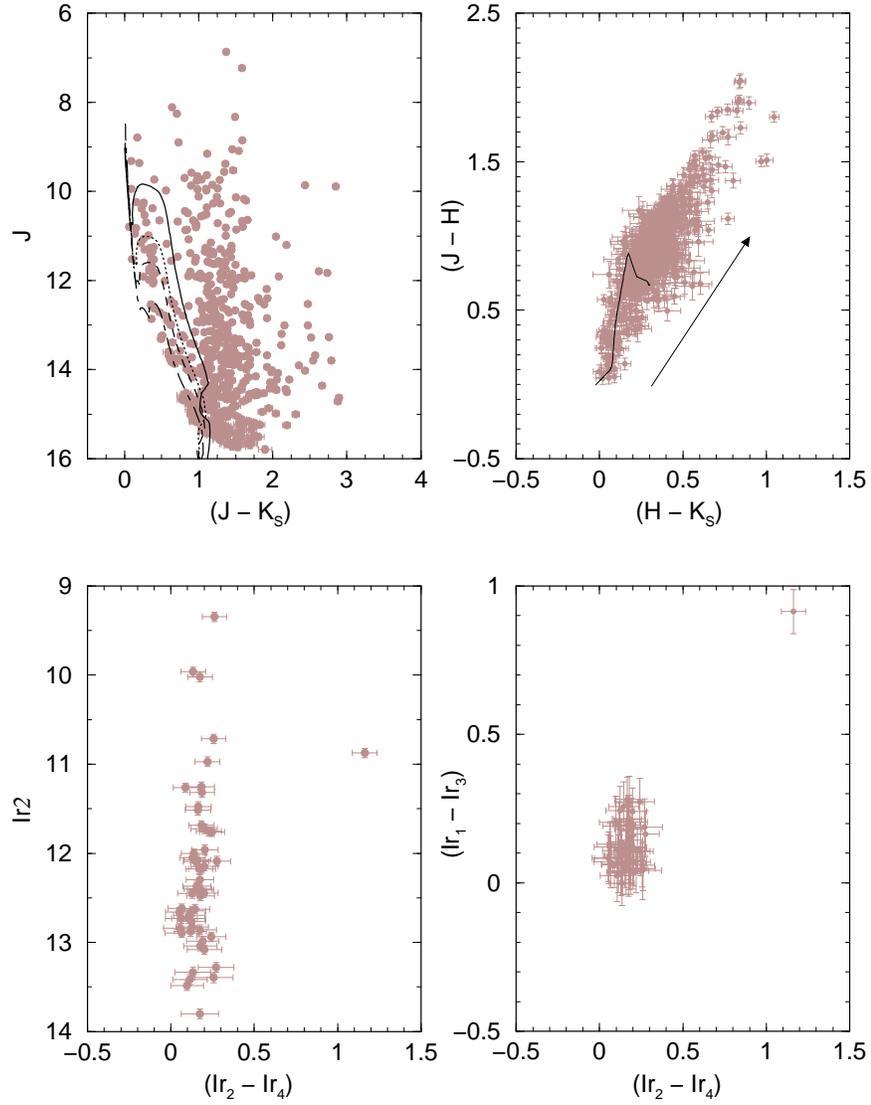}}
\caption{IC\,5146 Streamer : Decontaminated CMDs fitted when available with MS and PMS PADOVA isochrones. Top: 2MASS - $J\times(J-Ks)$ and $(J-H)\times(H-Ks)$. Bottom: Spitzer - $Ir2\times(Ir2-Ir4)$ and $(Ir1-Ir3)\times (Ir2-Ir4)$}
\label{fig_str}
\end{figure}

The distribution of sources in each EC is strikingly similar, but  some distinct concentrations of stars are also apparent  (Figs. \ref{cmds} to \ref{fig_str}).

%REFAZER FIG. com CLASSES especificadas

All Ecs in the sample have central density excesses (Fig. \ref{rdps}). The ECs appear to have an extra dust content, possibly related to a younger age. On the other hand, IC\, 5146 has a prominent RDP core. The RDPs in Fig. 5 were constructed with Spitzer data, except for NBB\,1 that is not in the Spitzer field, so its RDP was built with WISE data.

\section{Concluding remarks}
\label{conclu}

Ages throughout the complex are approximately constant at $t \sim 5$ Myr (Table \ref{tab2}). V absorptions are strong. IC\,5146 and the Streamer characterize the complex, at distance $1.2 - 1.3$ kpc, NBB\,2 , NBB\,3 and NBB\,4 appear to lie a bit farther (Table \ref{tab2}).

The Streamer is an optical dark nebula (Fig.1), that in the near IR is composed of many stars. The Streamer is a prestellar region (Roy et al. 2011, Kramer et al. 2003) with essencially no emission from warm grains. It  might be a particular stage in evolution of clouds, with a mix dust phases at different temperatures. Dobashi et al. (2003) have suggested that outflows have played an important role in supporting the parent cloud from collapsing. We conclude that the Streamer is stiil a \textit{Pandora} box to be further explored.

The present analyses suggest that the IC\,5146 cluster, the Streamer and the 5 newly found embedded clusters are neighbouring primeval examples of extended star formation and they are crucial for future more detailed studies of IC\,5146 complex, and \textbf{for comparisons with other star forming regions.}

\section*{Acknowledgements}
We thank Paul Harvey, the University of Texas at Austin and The GO-$4$ Spitzer Legacy project for providing us with Spitzer data. We make use of the 2MASS and WISE databases. We thank an anonymous referee for important remarks.  We acknowledge financial support from the Brazilian Institution CNPq.

%thank an anonymous referee for important comments and suggestions.  


\section*{References}
 \label{thebibliography}

 \bibliographystyle{elsarticle-harv} 
% \bibliography{<your bibdatabase>}


\end{document}